\documentclass[12pt,reqno]{amsart}

\usepackage{amsaddr}
\usepackage{graphicx} 

\newtheorem{claim}{Claim}[section]
\newtheorem{theorem}[claim]{Theorem}

\newtheorem{remark}[claim]{Remark}

\newtheorem{definition}[claim]{Definition}

\numberwithin{equation}{section}

\newcommand{\bo}{{\rm O}}

\newcommand{\ds}{\displaystyle}
\newcommand{\dsum}{\ds\sum}

\begin{document}
\title[Determinant for the damped wave equation]{Spectral determinant for the damped wave equation on an interval}
\author{Pedro Freitas} 
\address{Departamento de Matem\'atica, Instituto Superior T\'ecnico, Universidade de Lisboa, Av. Rovisco Pais 1,
P-1049-001 Lisboa, Portugal {\rm and}
Grupo de F\'isica Matem\'{a}tica, Faculdade de Ci\^encias, Universidade de Lisboa,
Campo Grande, Edif\'icio C6, P-1749-016 Lisboa, Portugal}
\email{psfreitas@fc.ul.pt}

\author{Ji\v{r}\'{\i} Lipovsk\'{y}}
\address{Department of Physics, Faculty of Science, University of Hradec Kr\'alov\'e, Rokitansk\'eho 62,
500\,03 Hradec Kr\'alov\'e, Czechia}
\email{jiri.lipovsky@uhk.cz}

\begin{abstract}
We evaluate the spectral determinant for the damped wave equation on an interval of length~$T$ with Dirichlet boundary conditions,
proving that it does not depend on the damping. This is achieved by analysing the square of the damped wave operator
using the general result by Burghelea, Friedlander, and Kappeler on the determinant for a differential operator with matrix coefficients.
\end{abstract}

\maketitle


\section{Introduction}
We consider the simple mathematical model of wave propagation on a damped string fixed at both ends given by
\begin{equation}\label{dwave}
  \frac{\partial^2 v(t,x)}{\partial t^2} + 2 a(x) \frac{\partial v(t,x)}{\partial t} = \frac{\partial^2 v(t,x)}{\partial x^2},
\end{equation}
with the space variable $x$ on an interval $[0,T]$, $v(0)=v(T)=0$ and $a(x) \in C([0,T])$. 
Despite its apparent simplicity, the problem is nontrivial and interesting and has received much attention over the last two
decades -- see, for instance,~\cite{GH,FL,CFNS,CZ, BoF}.

The operator associated with~\eqref{dwave} is non-selfadjoint and the asymptotical location of its eigenvalues was
determined to first order in~\cite{CFNS,CZ}, where it was shown that eigenvalues $\lambda$ converge to the vertical
line $\mathrm{Re\,}\lambda = -\left<a\right>$ as their imaginary part goes to $\pm \infty$, where
$\left<a\right>$ denotes the average of the damping function. The general asymptotic behaviour was analysed in~\cite{BoF},
where further spectral invariants were determined. 

Coming from other sources in the literature, the notion of determinant of a matrix has been generalised to operators.
In this analogy, we would like to obtain a regularisation corresponding to the product of eigenvalues of the given operator.
If the considered operator $\mathcal{S}$ has eigenvalues $\left\{\lambda_j\right\}_{j=1}^\infty$, in agreement with~\cite{rasi} (see
also~\cite{geya,MP}) we define the generalised zeta function associated with the operator $\mathcal{S}$ by
$$ 
\zeta_{\mathcal{S}}(s) = \sum_{j=1}^\infty \lambda_j^{-s},
$$
for complex $s$ in a half-plane such that the above Dirichlet series converges. The spectral determinant may then be
defined by the formula
\begin{equation}
  \mathrm{Det\,}\mathcal{S} = \mathrm{e}^{-\zeta_{\mathcal{S}}'(0)}\,,\label{eq:det}
\end{equation}
where the prime denotes the derivative with respect to the variable $s$. Note that the series defining the zeta function will not,
in general, be convergent for $s=0$. We use the definition of $\zeta_{\mathcal{S}}$ for the real part of $s$ large enough and understand the formula in the sense of the analytic continuation of the generalized zeta function to the complex plane. 

The spectral determinant was computed for the Sturm-Liouville operator in~\cite{LS}, where an elegant expression using the solution of a corresponding Cauchy problem was presented. This was extended to the case of quantum graphs in~\cite{ACDMT,frie}. We point out
that spectral determinants have several applications e.g. in quantum field theory \cite{Dunne}.

To the best of our knowledge the spectral determinant for the damped wave equation had not been studied previously, so in this note 
we bring together these two topics and evaluate this object. From a mathematical perspecive there is also what we believe to be
the interesting feature of applying the concept of the determinant of an operator to a non-selfadjoint operator. Furthermore, and
as we will see, the determinant does not, in fact, depend on the damping. This may be expected form a formal analysis, and our purpose
is to give a rigorous justification of this fact.

This note is structured as follows. In the next section, we ellaborate on the mathematical description of the model and state
the main result. In Section~\ref{sec:nodam} we then address the problem for the case without damping, as this already displays
some of the important features which we will need to consider later, namely, the fact that the associated zeta function will
depend on the branch cut which is chosen for the logarithm. In Section~\ref{sec:gen} we recall the general result of Burghelea,
Friedlander, and Kappeler. In Section~\ref{sec:square} we apply this result to the square of our operator, since a direct
application is not possible. Finally, we find the sought determinant for the damped wave equation in Section~\ref{sec:h}.

\section{Basic setting and formulation of the main result}
 Equation~\eqref{dwave} may be written in a different form, namely,
$$
  \frac{\partial}{\partial t} \begin{pmatrix}v_0(t,x)\\v_1(t,x)\end{pmatrix} = \begin{pmatrix}0 & 1 \\ \frac{\partial^2}{\partial x^2}& -2 a(x)\end{pmatrix}\begin{pmatrix}v_0(t,x)\\v_1(t,x)\end{pmatrix}
$$
which will prove to be more convenient for our purposes.
Using the ansatz $v_0 (t,x) = \mathrm{e}^{\lambda t} u_0(x)$, $v_1 (t,x) = \mathrm{e}^{\lambda t} u_1(x)$, we can translate the
initial value problem into the following spectral problem 
$$
  \mathcal{H} \begin{pmatrix}u_0(x)\\ u_1(x)\end{pmatrix} = \lambda \begin{pmatrix}u_0(x)\\ u_1(x)\end{pmatrix} 
$$
where $\mathcal{H}$ denotes the matrix operator 
$$
  \mathcal{H} = \begin{pmatrix}0 & 1 \\ \frac{\partial^2}{\partial x^2}& -2 a(x)\end{pmatrix}\,.
$$
The domain of this operator consists of functions $\mathbf{u}(x) = \begin{pmatrix}u_0(x)\\ u_1(x)\end{pmatrix}$ with components in the Sobolev spaces $u_j(x) \in W^{2,2}([0,T])$, $j = 0,1$ satisfying the Dirichlet boundary conditions
$$
  u_j (0) = u_j(T) = 0\,,\quad j = 0,1\,.
$$

Our main result is the following.
\begin{theorem}\label{thm:main}
Assume $a(x)\in C([0,T])$, and let $\varepsilon$ be a positive number such that there are no eigenvalues with phase on the interval
$[\pi-\varepsilon,\pi)$, Then the spectral determinant of the operator $\mathcal{H}$ does not depend on the damping and equals
$\pm 2T$, where the plus and minus signs correspond to whether we define $\lambda_j^{-s}=\mathrm{e}^{-s\log{\lambda_j}}$ in such a way that the branch cut of the logarithm is $\lambda = t \mathrm{e}^{i(\pi-\varepsilon)}$, or $\lambda = t \mathrm{e}^{i(2\pi-
\varepsilon)}$, $t\in [0,\infty)$, respectively.
\end{theorem}

\begin{remark}
Note that a value of $\varepsilon$ as above always exists, since on any compact set there are only a finite number of eigenvalues.
\end{remark}

\begin{remark}
The case of the damped wave equation where a potential is added to the right-hand side of~\eqref{dwave} may be treated in a similar fashion and the corresponding determinant also turns out to be independent of the damping term. We discuss this situation in
Remark~\ref{rem:pot}.
\end{remark}


\section{The case of $a(x)=0$}\label{sec:nodam}
We begin by considering the case without damping, and denote the corresponding operator by $\mathcal{H}_0$. It is a simple exercise that its eigenvalues are of the form $\lambda_j = \frac{i j \pi}{T}$, $j\in \mathbb{Z}\backslash \{0\}$. To obtain the spectral determinant in this instance, we start from the zeta function resulting from the definition~\eqref{eq:det}. However, one must
proceed carefully here, as the result depends on the definition of $\lambda_j^{-s}$ and, in particular, on which branch of
the logarithm we use when defining $i^{-s}$ and $(-i)^{-s}$. 

First, we consider that the logarithm has the cut in the negative real axis, i.e. the eigenvalues of $\mathcal{H}_0$ in the upper half-plane are $\lambda_j = \frac{j\pi}{T}\mathrm{e}^{\frac{i\pi}{2}}$, $j\in \mathbb{N}$ and the eigenvalues in the lower half-plane are $\lambda_j = \frac{j\pi}{T}\mathrm{e}^{-\frac{i\pi}{2}}$, $j\in \mathbb{N}$.

The generalized zeta function for this operator is 
\begin{eqnarray*}
  \zeta_{\mathcal{H}_0}(s) & = & \sum_{j=1}^\infty \left[\left(\frac{j\pi}{T}\mathrm{e}^{\frac{i\pi}{2}}\right)^{-s}+\left(\frac{j\pi}{T}\mathrm{e}^{-\frac{i\pi}{2}}\right)^{-s}\right]
\\
 & = & \sum_{j=1}^\infty \left(\mathrm{e}^{-\frac{i\pi}{2}s}+\mathrm{e}^{\frac{i\pi}{2}s}\right) \left(\frac{j\pi}{T}\right)^{-s} 
\\
 & = & 2\mathrm{e}^{s\log{\frac{T}{\pi}}}\cos{\left(\frac{\pi s}{2}\right)}\zeta_\mathrm{R}(s)\,,
\end{eqnarray*}
where $\zeta_\mathrm{R}(s) = \dsum_{j=1}^\infty j^{-s}$ is the Riemann zeta function. We obtain
$$
  -\zeta_{\mathcal{H}_0}'(0) =-2\log{\frac{T}{\pi}} \zeta_{\mathrm{R}}(0)-2\zeta_{\mathrm{R}}'(0) = \log{\frac{T}{\pi}}+\log{(2\pi)} = \log{(2T)}\,,
$$
where we have used $\zeta_{\mathrm{R}}(0) = -\frac{1}{2}$ and $\zeta_{\mathrm{R}}'(0) = -\frac{1}{2}\log{(2\pi)}$. Hence the spectral 
determinant for the operator $\mathcal{H}_0$ is given by
$$
  \mathrm{Det\,}\mathcal{H}_0 = \mathrm{e}^{-\zeta_{\mathcal{H}_0}'(0)} = 2T\,.
$$

Now we are going to compute the determinant in the case where we choose the cut to be the positive real axis. The eigenvalues are $\lambda_j = \frac{j\pi}{T}\mathrm{e}^{\frac{i\pi}{2}}$, $j\in \mathbb{N}$, for the upper half-plane and $\lambda_j = \frac{j\pi}{T}\mathrm{e}^{\frac{3i\pi}{2}}$, $j\in \mathbb{N}$ for the lower half-plane. The generalized zeta function is now
\begin{eqnarray*}
  \zeta_{\mathcal{H}_0}(s) & = & \sum_{j=1}^\infty \left[\left(\frac{j\pi}{T}\mathrm{e}^{\frac{i\pi}{2}}\right)^{-s}+\left(\frac{j\pi}{T}\mathrm{e}^{\frac{3i\pi}{2}}\right)^{-s}\right]
\\
 & = & \mathrm{e}^{-i\pi s}\sum_{j=1}^\infty \left(\mathrm{e}^{\frac{i\pi}{2}s}+\mathrm{e}^{-\frac{i\pi}{2}s}\right) \left(\frac{j\pi}{T}\right)^{-s} 
\\
 & = & 2\mathrm{e}^{-i\pi s}\mathrm{e}^{s\log{\frac{T}{\pi}}}\cos{\left(\frac{\pi s}{2}\right)}\zeta_\mathrm{R}(s)\,.
\end{eqnarray*}
Hence we have
$$
  -\zeta_{\mathcal{H}_0}'(0) =-2\log{\frac{T}{\pi}} \zeta_{\mathrm{R}}(0)+2i\pi \zeta_\mathrm{R}(0) -2\zeta_{\mathrm{R}}'(0) = \log{\frac{T}{\pi}}-i \pi + \log{(2\pi)} = -i\pi+ \log{(2T)}\,.
$$
The spectral determinant for the operator $\mathcal{H}_0$ is 
$$
  \mathrm{Det\,}\mathcal{H}_0 = \mathrm{e}^{-\zeta_{\mathcal{H}_0}'(0)} = -2T\,.
$$

\section{A general result}\label{sec:gen}
The starting point for finding the determinant for the damped wave equation is a general result by Burghelea, Friedlander and Kappeler \cite{BFK}. This result gives a formula for the determinant of a more general matrix-valued operator on an interval. For convenience,
we state this result here, together with the necessary definitions.

\begin{definition}\label{def:bfk}
Let us for $n\in\mathbb{N}$ define the operator $\mathcal{A} = \dsum_{k = 0}^{2n} a_k(x)(-i)^k \frac{\mathrm{d}^k}{\mathrm{d}x^k}$, where $a_k$ are $r\times r$ matrices, in general smoothly dependent on $x\in [0,T]$. We assume that the leading term $a_{2n}$ is nonsingular and that there exist an angle $\theta$ so that $\mathrm{spec}\,a_{2n}\cap \{\rho\mathrm{e}^{i\theta}, 0\leq \rho <\infty\} = \emptyset$. We assume the following boundary conditions at the end points of the interval.
$$
  \sum_{k = 0}^{\alpha_j} b_{jk}u^{(k)}(T) = 0\,,\quad \sum_{k = 0}^{\beta_j} c_{jk}u^{(k)}(0) = 0\,,\quad 1\leq j \leq n	\,.
$$
Here, $b_{jk}$ and $c_{jk}$ are for each $j,k$ constant $r\times r$ matrices and $b_{j\alpha_j} = c_{j\beta_j} = I$ ($I$ denotes
the $r\times r$ identity matrix). The integer numbers $\alpha_j$ and $\beta_j$ satistfy
\begin{eqnarray*}
  0\leq \alpha_1 <\alpha_2 <\dots<\alpha_n<2n-1\,,\\
  0\leq \beta_1 <\beta_2 <\dots<\beta_n<2n-1\,.
\end{eqnarray*}

Moreover, we define $|\alpha| = \sum_{j = 1}^n \alpha_j$, $|\beta| = \sum_{j = 1}^n \beta_j$. We define the $2n\times 2n$ matrices $B= (B_{jk})$ and $C= (C_{jk})$, whose entries are $r\times r$ matrices. Here $1\leq j\leq 2n$, $0\leq k \leq 2n-1$ and 
\begin{eqnarray*}
  B_{jk} &:=&\left\{\begin{matrix}b_{jk} & \mathrm{for\quad}1\leq j\leq n \quad\mathrm{and\quad} 0\leq k \leq \alpha_j \\
								0      & \mathrm{otherwise}\end{matrix} \right.\,,\\
  C_{jk} &:=&\left\{\begin{matrix}b_{j-n,k} & \mathrm{for\quad}n+1\leq j\leq 2n \quad\mathrm{and\quad} 0\leq k \leq \alpha_{j-n} \\
								0      & \mathrm{otherwise}\end{matrix} \right.\,.
\end{eqnarray*}

We define a $2n\times 2n$ matrix $Y(x) = (y_{k\ell}(x))$, $0\leq k, \ell\leq 2n-1$ whose entries are $r\times r$ matrices
$y_{k\ell} (x)$ defined by 
$$
  y_{k\ell}(x) := \frac{\mathrm{d}^{k}y_\ell(x)}{\mathrm{d}x^{k}}\,,
$$
where $y_{\ell}(x)$ is the solution of the Cauchy problem $\mathcal{A} y_\ell(x) = 0$ with the initial conditions $y_{k\ell}(0) = \delta_{k\ell} I$. We are interested in the value of the matrix $Y$ at the point $T$.

Finally, we introduce 
\begin{eqnarray*}
  g_\alpha &:=& \frac{1}{2}\left(\frac{|\alpha|}{n}-n+\frac{1}{2}\right)\,,\\
  h_\alpha &:=& \mathrm{det\,}\begin{pmatrix}w_1^{\alpha_1} & \dots & w_n^{\alpha_1}\\ \vdots & \ddots & \vdots \\ w_1^{\alpha_n} & \dots & w_n^{\alpha_n}\\ \end{pmatrix}\,,
\end{eqnarray*}
where $w_k = \mathrm{exp\,}\left(\frac{2k-n-1}{2n}\pi i\right)$. Similarly, we define $g_\beta$ and $h_\beta$. We denote by $\gamma_j$, $j = 1,\dots, r$ the eigenvalues of the matrix $a_{2n}$ and define
$$
  {(\mathrm{det\,}a_{2n})}_\theta^{g_{\alpha}} :=\prod_{j = 1}^r |\gamma_j|^{g_\alpha} \mathrm{exp\,}(ig_\alpha \mathrm{arg}(\gamma_j)) 
$$
with $\theta -2\pi<\mathrm{arg\,}\gamma_j <\theta$.
\end{definition}

\begin{theorem}\label{thm:bfk}(Burghelea, Friedlander and Kappeler)\\
The spectral determinant for the operator $\mathcal{A}$ is 
$$
  \mathrm{Det\,}\mathcal{A} = K_\theta\mathrm{exp\,}\left(\frac{i}{2}\int_0^T \mathrm{Tr\,}(a_{2n}(x)^{-1}a_{2n-1}(x))\,\mathrm{d}x\right) \,\mathrm{det}(BY(T)-C)\,,
$$
where 
$$ 
  K_\theta = [(-1)^{|\beta|}(2n)^{n}h_\alpha^{-1}h_\beta^{-1}]^r  {(\mathrm{det\,}a_{2n}(0))}_\theta^{g_{\beta}}  {(\mathrm{det\,}a_{2n}(T))}_\theta^{g_{\alpha}}\,.
$$

\end{theorem}

\section{The square of the operator $\mathcal{H}$}\label{sec:square}
Our purpose is to use Theorem~\ref{thm:bfk} to obtain the spectral determinant for the operator $\mathcal{H}$. However, a direct 
application of the theorem is not possible since the highest order derivative is only present in one of the entries of the matrix 
which gives the operator $\mathcal{H}$. This would contradict the assumption that the matrix $a_{2n}$ is nonsingular. In order
to overcome this difficuly, we shall consider the operator $\mathcal{A} = \mathcal{H}\circ\mathcal{H}=\mathcal{H}^2$ for which it is then possible to apply
Theorem~\ref{thm:bfk}. 

However, we have to be careful when computing the spectral determinant of $\mathcal{H}$ from the spectral determinant of $\mathcal{A}$,
as the determinant of the composition of two operators does not necessarily have to equal the product of their determinants. This is known
in the literature as the  \emph{multiplicative anomaly} and, as has been shown in~\cite{BS,Woj}, even the determinant of the square of an
operator is not always the square of the determinant.



A direct calculation yields
\begin{eqnarray*}
  \mathcal{A} & = & \begin{pmatrix}0 & 1 \\ \frac{\partial^2}{\partial x^2}& -2 a(x)\end{pmatrix}^2 = \begin{pmatrix}\frac{\partial^2}{\partial x^2} & -2a(x)\\ -2a(x)\frac{\partial^2}{\partial x^2} & \frac{\partial^2}{\partial x^2}+4a^2(x)\end{pmatrix}
\\
 & = & \begin{pmatrix}-1 & 0\\ 2a(x)& -1\end{pmatrix}\left(-i\frac{\partial}{\partial x}\right)^2 + \begin{pmatrix}0 & -2a(x)\\ 0 & 4a^2(x)\end{pmatrix} \left(-i\frac{\partial}{\partial x}\right)^0\,.
\end{eqnarray*}
Hence we have $n = 1$,
$$
  a_{2}(x) =\begin{pmatrix}-1 & 0\\ 2a(x)& -1\end{pmatrix}\,,\quad a_1 = \begin{pmatrix}0 & 0 \\ 0 & 0 \end{pmatrix}\,,\quad a_0 = \begin{pmatrix}0 & -2a(x)\\ 0 & 4a^2(x)\end{pmatrix} \,.
$$
We deal with matrices $2\times 2$, so $r=2$, and have $\alpha_1 = \beta_1 = 0$ and hence $|\alpha| = |\beta| = 0$. 

The boundary conditions according to Definition~\ref{def:bfk} are
$$
  b_{10}\mathbf{u}(T)+b_{11}\mathbf{u}'(T) = 0\,,\quad c_{10}\mathbf{u}(0)+c_{11}\mathbf{u}'(0) = 0\,.
$$
We have the Dirichlet boundary conditions $\mathbf{u}(T) = \mathbf{u}(0) = \mathbf{0}$, and thus the matrices $b_{jk}$ and
$c_{jk}$ must be chosen as
$$
  b_{10} = c_{10} = I\,,\quad b_{11} = c_{11} = 0\,,
$$
where $I$ is the $2\times 2$ identity matrix.

The matrices $B$ and $C$ are $2\times 2$ matrices with the entries being $2\times 2$ matrices. Their rows are indexed by 1 and 2, their columns by 0 and 1. According to Definition~\ref{def:bfk} we have
\begin{eqnarray*}
  B_{10} = b_{10} = I\,,\quad B_{11} = B_{20} = B_{21} = 0\,,\\
  C_{20} = c_{10} = I\,,\quad C_{11} = C_{10} = C_{21} = 0\,.\\
\end{eqnarray*}
Hence we have
$$
  B = \begin{pmatrix}I & 0 \\ 0 & 0\end{pmatrix}\,,\quad C = \begin{pmatrix}0 & 0 \\ I & 0\end{pmatrix}\,.
$$

The matrix $a_{2n} = a_2$ clearly has eigenvalues $\gamma_1 = \gamma_2 = -1$. Moreover, we have
$$
  g_\alpha = g_\beta = \frac{1}{2}\left(\frac{0}{2}-1+\frac{1}{2}\right) = -\frac{1}{4}\,,\quad h_\alpha = h_\beta = w_1^0 = 1\,.
$$
and
$$
  (\mathrm{det\,}a_{2n})_\theta^{g_\alpha} = \prod_{j=1}^2 1^{-1/4}\mathrm{e}^{i(-1/4)(-\pi)} = i
$$
To sum up,
$$
  K_\theta = [(-1)^0(2\cdot 1)^1 1^{-2}]^2 i^2 = -4\,.
$$

Now, we are going to compute the matrix $Y(x)$. By its definition, we have
$$
  Y(x) = \begin{pmatrix}y_{00}(x) & y_{01}(x)\\ y_{10}(x) & y_{11}(x)\end{pmatrix} = \begin{pmatrix} y_0(x) & y_1(x)\\ y_0'(x) & y_1'(x)\end{pmatrix}\,,
$$
where the entries of this matrix are $2\times 2$ matrices with the following boundary conditions at $x=0$
\begin{equation}
  y_0(0) = I\,,\quad y_0'(0) = 0\,,\quad y_1(0) = 0\,,\quad y_1'(0) = I\,.\label{eq:bc}
\end{equation}
For the matrix in the formula in Theorem~\ref{thm:bfk} we have
\begin{eqnarray*}
  \mathrm{det\,}(BY(T)-C) & = & \mathrm{det\,}\left[\begin{pmatrix}I & 0\\ 0 & 0\end{pmatrix}\begin{pmatrix}y_0(T) & y_1(T)\\ y_0'(T)& y_1'(T)\end{pmatrix}-\begin{pmatrix}0 & 0 \\ I &0\end{pmatrix}\right] 
  \\
  & = & \mathrm{det\,}\begin{pmatrix}y_0(T) & y_1(T)\\ -I & 0\end{pmatrix} 
\\
  & = & \mathrm{det\,}y_1(T)\,.
\end{eqnarray*}

Now, we will find the solutions of the Cauchy problem, i.e. the equation
$$
  \mathcal{H}^2 \begin{pmatrix}u_0(x)\\u_1(x)\end{pmatrix} = 0
$$
with the boundary conditions~\eqref{eq:bc}. We obtain
\begin{eqnarray}
  u_0''(x)-2a(x)u_1(x) & = & 0\,,\label{eq:cauchy1}\\
  -2a(x)u_0''(x)+u_1''(x)+4a^2(x)u_1(x) & = & 0\,.\label{eq:cauchy2}
\end{eqnarray}
Multiplying~\eqref{eq:cauchy1} by $2a(x)$ and adding the result to~\eqref{eq:cauchy2} we obtain
\begin{equation}
  u_1''(x) = 0\,.\label{eq:cauchy3}
\end{equation}
The boundary conditions are $y^{(k)}_l(0) = \delta_{kl}I$. Hence to obtain the matrix $y_1(x)$, we have to find two vectors $\begin{pmatrix}u_{01}(x)\\u_{11}(x)\end{pmatrix}$ and $\begin{pmatrix}u_{02}(x)\\u_{12}(x)\end{pmatrix}$, for which
\begin{equation}
  y_1(0) = \begin{pmatrix}u_{01}(0) & u_{02}(0)\\ u_{11}(0) & u_{12}(0)\end{pmatrix} = 0\,,\quad y_1'(0) = \begin{pmatrix}u_{01}'(0) & u_{02}'(0)\\ u_{11}'(0) & u_{12}'(0)\end{pmatrix}= I\,.\label{eq:cauchy4}
\end{equation}
The general form of the solution of \eqref{eq:cauchy3} is $u_1(x) = ax+b$, and from the first condition in \eqref{eq:cauchy4} we have $b=0$. The second condition in \eqref{eq:cauchy4} yields
$$
  u_{11}(x) = 0\,,\quad u_{12}(x) = x\,.
$$

Hence in the first case we have from \eqref{eq:cauchy1} $u_{01}''(x) = 0$, hence we have $u_{01}(x) = cx+d$. By the first condition in \eqref{eq:cauchy4}  we have $d=0$, by the second one $u_{01}(x) = x$.

In the second case ($u_{12}(x) = x$) we obtain from \eqref{eq:cauchy1} $u_{02}''(x) = 2a(x)x$. Hence with the use of the second condition in \eqref{eq:cauchy4} we have $u_{02}'(x) = \int_0^x 2a(s)s\,\mathrm{d}s$ and with the use of the first condition in \eqref{eq:cauchy4} we have $u_{02}(x) = \int_0^x \int_0^q 2a(s)s\,\mathrm{d}s\mathrm{d}q$.

To sum up,
$$
  y_1(x) = \begin{pmatrix}x &  \int_0^x \int_0^q 2a(s)s\,\mathrm{d}s\mathrm{d}q\\ 0 & x\end{pmatrix}
$$
and hence $\mathrm{det\,}y_1(T) = T^2$. We obtained 
$$
  \mathrm{Det}\,\mathcal{A} = -4T^2\,.
$$

\section{The determinant of $\mathcal{H}$} \label{sec:h}
Let us consider the operator $\mathcal{H}$ in the general case. In each horizontal strip $|\mathrm{Im\,}\lambda|<K$ there are finitely many eigenvalues. Hence there exists a positive $\varepsilon$ so that there are no eigenvalues whose arguments are in the intervals $(0,\varepsilon]$, $[\pi-\varepsilon,\pi)$, $(\pi,\pi+\varepsilon]$, and  $[2\pi-\varepsilon,2\pi)$. We will choose the cut of the logarithm as the half-line $\lambda = t \mathrm{e}^{i(\pi-\varepsilon)}$, $t\in [0,\infty)$. We denote the eigenvalues of $\mathcal{H}$ in the upper half-plane by $\tilde\mu_j$, and their complex conjugates by $\bar{\tilde\mu}_j$, the positive real eigenvalues by $\tilde\nu_j$, $j\in I_1$ (here $I_1$ is a finite index set) and the negative real eigenvalues as $-\tilde\omega_j$, $j\in I_2$ (here $I_2$ is a finite index set). It can be easily proven that there are no zero eigenvalues of $\mathcal{H}$. 

For the operator $\mathcal{A}$, we denote the eigenvalues as $\mu_j^2$, $\bar{\mu}_j^2$, $\nu_j^2$ and $\omega_j^2$. We consider the phases of all these eigenvalues  in the interval $[0,2\pi-2\varepsilon)$, the cut of the logarithm for the operator $\mathcal{A}$ is the half-line $\lambda = t \mathrm{e}^{-2i\varepsilon}$. Hence one can easily obtain
\begin{eqnarray*}
  \log{\tilde \mu_j} = \frac{1}{2}\log{\mu_j^2}\,,\quad \log{\bar{\tilde\mu}_j} = \frac{1}{2}\log{\bar{\mu}_j^2} - \pi i\,,\\
  \log{\tilde \nu_j} = \frac{1}{2}\log{\nu_j^2}\,,\quad \log{(-\tilde\omega_j)} = \frac{1}{2}\log{\omega_j^2} - \pi i\,.\\
\end{eqnarray*}

The zeta functions for these operators are
\begin{eqnarray*}
    \zeta_{\mathcal{A}}(s) & = & \sum_{j=1}^\infty [({\mu}_j^2)^{-s}+({\bar{\mu}_j}^2)^{-s}]+\sum_{j\in I_1}({\nu}_j^2)^{-s}+\sum_{j\in I_2}(({-\omega}_j)^2)^{-s}
\\
   & = & \sum_{j=1}^\infty (\mathrm{e}^{-s\log{\mu_j^2}}+\mathrm{e}^{-s\log{\bar{\mu}_j^2}})+\sum_{j\in I_1}\mathrm{e}^{-s\log{\nu_j^2}}+\sum_{j\in I_2}\mathrm{e}^{-s\log{\omega_j^2}} \,.
\\
  \zeta_{\mathcal{H}}(s) & = & \sum_{j=1}^\infty [{\tilde \mu}_j^{-s}+{\bar{\tilde \mu}_j}^{-s}]+\sum_{j\in I_1}{\tilde \nu}_j^{-s}+\sum_{j\in I_2}({-\tilde \omega}_j)^{-s}
\\
  & = & \sum_{j=1}^\infty (\mathrm{e}^{-s\log{{\tilde\mu}_j}}+\mathrm{e}^{-s\log{\bar{\tilde\mu}_j}})+\sum_{j\in I_1}\mathrm{e}^{-s\log{{\tilde\nu}_j}}+\sum_{j\in I_2}\mathrm{e}^{-s\log{(-\tilde\omega_j)}}
\\
 & = & \sum_{j=1}^\infty (\mathrm{e}^{-\frac{1}{2}s\log{\mu_j^2}}+\mathrm{e}^{-\frac{1}{2}s\log{\bar{\mu}_j^2}}\mathrm{e}^{\pi i s})+\sum_{j\in I_1}\mathrm{e}^{-\frac{1}{2}s\log{\nu_j^2}}+\sum_{j\in I_2}\mathrm{e}^{-\frac{1}{2}s\log{\omega_j^2}}\mathrm{e}^{\pi i s}\,.
\end{eqnarray*}
Hence we have
\begin{eqnarray*}
  \zeta_{\mathcal{H}}(s) - \zeta_{\mathcal{A}}\left(\frac{s}{2}\right) & = & (\mathrm{e}^{\pi i s}-1) \left(\sum_{j=1}^\infty \mathrm{e}^{-\frac{1}{2}s\log{\bar{\mu}_j^2}}+ \sum_{j\in I_2}\mathrm{e}^{-\frac{1}{2}s\log{\omega_j^2}}\right)
\\
& = & 2i\mathrm{e}^{\frac{\pi i s}{2}} \sin{\frac{\pi s}{2}}\left(\sum_{j=1}^\infty \mathrm{e}^{-\frac{1}{2}s\log{\bar{\mu}_j^2}}+ \sum_{j\in I_2}\mathrm{e}^{-\frac{1}{2}s\log{\omega_j^2}}\right)\,.
\end{eqnarray*}
For the derivatives at zero we obtain, using the fact that the sine will vanish in that case,
\begin{eqnarray*}
  \zeta_{\mathcal{H}}'(0) - \frac{1}{2}\zeta_{\mathcal{A}}'(0) & = &  i \pi \lim_{s\to 0} \sum_{j=1}^\infty \mathrm{e}^{-\frac{1}{2}s\log{\bar{\mu}_j^2}}+ i \pi\, \mathrm{card\,}I_2
\\
  & = & i \pi\, \mathrm{card\,}I_2 + i \pi \lim_{s\to 0} \sum_{j=1}^\infty \mathrm{e}^{-\frac{i\pi s}{2}}\mathrm{e}^{s\log{\frac{T}{\pi}}}\mathrm{e}^{-s\log{j}}
\\
  & & \hspace{1cm} + i \pi \lim_{s\to 0} \sum_{j=1}^\infty\left(\mathrm{e}^{-\frac{1}{2}s\log{\bar{\mu}_j^2}} - \mathrm{e}^{-\frac{i\pi s}{2}}\mathrm{e}^{s\log{\frac{T}{\pi}}}\mathrm{e}^{-s\log{j}}\right)\,.
\end{eqnarray*}

We will prove that the last term is zero. We will use the asymptotics (see e.g. \cite{BoF})
$$
  \bar{\mu}_j^2 = -\frac{j^2\pi^2}{T^2}\left(1+\bo\left(\frac{1}{j}\right)\right)
$$
and by our assumption $\bar{\mu}_j^2$ is not close to zero. We have
\begin{eqnarray*}
  \lim_{s\to 0} \left|\sum_{j=1}^\infty (\mathrm{e}^{-\frac{1}{2}s\log{\bar{\mu}_j^2}}- \mathrm{e}^{-\frac{i\pi s}{2}}\mathrm{e}^{s\log{\frac{T}{\pi}}}\mathrm{e}^{-s\log{j}})\right|
 & = & \lim_{s\to 0} \left|\sum_{j=1}^\infty \mathrm{e}^{-\frac{i\pi s}{2}}\mathrm{e}^{s\log{\frac{T}{\pi}}}\mathrm{e}^{-s\log{j}}\right.
\\ 
 && \hspace{1cm} \left. \times (\mathrm{e}^{-\frac{1}{2}s\log{(1+\bo(1/j))}}-1)\right|
\\
 & \leq & \lim_{s\to 0} \left|\sum_{j=1}^\infty \mathrm{e}^{-s\log{j}}\right| |\mathrm{e}^{s\log{\frac{T}{\pi}}}| |\mathrm{e}^{-sC}-1|
\\
 & = & \lim_{s\to 0} |\zeta_{\mathrm{R}}(s)| |\mathrm{e}^{s\log{\frac{T}{\pi}}}| |\mathrm{e}^{-sC}-1| 
\\
 & = & 0\,.
\end{eqnarray*}
where $C$ is a real constant and $\zeta_{\mathrm{R}}(0) = -\frac{1}{2}$. Hence we have
\begin{equation}
  \zeta_{\mathcal{H}}'(0) - \frac{1}{2}\zeta_{\mathcal{A}}'(0) = i \pi\, \mathrm{card\,}I_2 + i \pi \zeta_{\mathrm{R}}(0) =i \pi\, \mathrm{card\,}I_2 - \frac{i \pi}{2} \,.\label{eq:zetaprime}
\end{equation}
Then the determinant is equal to
$$
  \mathrm{Det\,}\mathcal{H} = \mathrm{e}^{-\zeta_{\mathcal{H}}'(0)} = \mathrm{e}^{-\frac{1}{2}\zeta_{\mathcal{A}}'(0)} \mathrm{e}^{\frac{i\pi}{2}}\mathrm{e}^{-i\pi \,\mathrm{card\,}I_2}= \mp i\sqrt{\mathrm{Det\,}\mathcal{A}}= \mp i\sqrt{-4T^2} = \pm 2T\,.
$$
The number of negative real eigenvalues $\mathrm{card\,}I_2$ is always even, so this term does not influence the result. The aim of the rest of our analysis is to find the sign of the determinant for the case of the cut we chose. First, we will follow an approach different from that which was used in Section~\ref{sec:nodam} to find the determinant for the operator without damping.
The zeta function for the operator~$\mathcal{A}_0$ is
$$
\zeta_{\mathcal{A}_0} (s)= 2 \sum_{j=1}^\infty \left(\frac{j\pi}{T}\right)^{-2s}(-1)^{-s} = 2 \left(\frac{T}{\pi}\right)^{2s}\mathrm{e}^{-s\log{(-1)}} \sum_{j=1}^\infty j^{-2s} 
= 2 \mathrm{e}^{-i\pi s} \left(\frac{T}{\pi}\right)^{2s} \zeta_{\mathrm{R}}(2s)\,.
$$
Then we have
$$
-\zeta_{\mathcal{A}_0}'(0) = 2\pi i \zeta_{\mathrm{R}}(0) - 4 \log{\frac{T}{\pi}}\zeta_{\mathrm{R}}(0)-4\zeta_{\mathrm{R}}'(0)=  -\pi i + 2 \log{(2T)}\,.
$$
Using equation~\eqref{eq:zetaprime} we obtain
$$
  \mathrm{Det\,}\mathcal{H}_0 = \mathrm{e}^{-\zeta_{\mathcal{H}_0'(0)}} = \mathrm{e}^{-¨\frac{1}{2}\zeta_{\mathcal{A}_0'(0)}} \mathrm{e}^{\frac{i\pi}{2}} = \mathrm{e}^{-\frac{i\pi}{2}} \mathrm{e}^{\frac{i\pi}{2}} \mathrm{e}^{\log{(2T)}} = 2T\,.
$$

Now, we are going to generalize this approach to the operator with damping. Our aim is to find the imaginary part of $\zeta_{\mathcal{H}_0}'(0)$. We denote the absolute value of $\mu_j^2$ by $r_j$ and its phase by $\pi -\varphi_j$. We have
$$
  \mu_j^2 = r_j\mathrm{e}^{i(\pi -\varphi_j)}\,,\quad \bar{\mu}_j^2 = r_j\mathrm{e}^{i(\pi +\varphi_j)}\,.
$$
We obtain
$$
  (\mu_j^2)^{-s}+ (\bar{\mu}_j^2)^{-s} = r_j^{-s}\mathrm{e}^{-i\pi s}(\mathrm{e}^{i\varphi_j s}+\mathrm{e}^{-i\varphi_j s}) = 2\mathrm{e}^{-i\pi s} r_j^{-s}\cos{(\varphi_j s)}
$$
and since $r_j^{-s}\cos{(\varphi_j s)}$, $\nu_j^2$ and $\omega_j^2$ are real
\begin{eqnarray*}
  \hspace{-5mm}\mathrm{Im\,}\frac{\mathrm{d}}{\mathrm{d}s}\left.\left\{\sum_{j=1}^\infty\left[(\mu_j^2)^{-s}+ (\bar{\mu}_j^2)^{-s}\right]+\sum_{j\in I_1}(\nu_j^2)^{-s}+\sum_{j\in I_1}(\omega_j^2)^{-s}\right\}\right|_{s=0} \!\! & = & \!\! -2\pi\lim_{s\to 0} \sum_{j=0}^\infty r_j^{-s}\cos{(\varphi_j s)} 
 \\
 \!\! & = & \!\! - \pi \lim_{s\to 0}\sum_{j=1}^\infty\left[(\mu_j^2)^{-s}+ (\bar{\mu}_j^2)^{-s}\right]
\end{eqnarray*}

We will use the asymptotic behaviour of the eigenvalues of the operator $\mathcal{H}$ (see \cite{BoF})
$$
  \mu_j = \frac{j\pi}{T}i-\left<a\right>-\frac{i \left<a^2\right>T}{2\pi j}+\bo\left(\frac{1}{j^2}\right)
$$
(here $\left<\cdot\right>$ denotes the average of the function on the interval) which leads to the following behaviour of the eigenvalues of the operator $\mathcal{A}$
$$
  \mu_j^2 = -\frac{j^2\pi^2}{T^2}-\frac{2j\pi \left<a\right>}{T}i +\bo(1)\,.
$$
Hence we obtain
$$
  (\mu_j^2)^{-s} = (-1)^{-s}j^{-2s} \frac{T^{2s}}{\pi^{2s}}\left(1-\frac{2T\left<a\right>s}{\pi j}i+\bo(j^{-2})\right)\,.
$$
Since $\bar{\mu}_j^2$ is the complex conjugate of $\mu_j^2$, we have
$$
  \sum_{j=1}^\infty \left[(\mu_j^2)^{-s}+(\bar{\mu}_j^2)^{-s}\right] = 2\frac{T^{2s}}{\pi^{2s}}(-1)^{-s} \sum_{j=0}^\infty j^{-2s} + \sum_{j=1}^\infty \bo(j^{-2s-2})\,.
$$
The second series is absolutely convergent down to ${\rm Re}\,s=0$, and hence we can exchange the sum and the limit. Moreover,
the term under the sum goes to zero as $s\to 0$, as it is multiplied by $s$. The first term can be written as $2\mathrm{e}^{-i\pi s}\frac{T^{2s}}{\pi^{2s}}\zeta_{\mathrm{R}}(2s)$, hence its limit is equal to $-1$. We conclude that
$$
  -\mathrm{Im\,}\zeta_{\mathrm{A}}'(0) = -\pi\,.
$$
Using equation~\eqref{eq:zetaprime} similarly to the case of no damping leads to the positive sign for the determinant with the cut of the logarithm taken just above the negative real axis. This proves Theorem~\ref{thm:main}. 

\begin{remark}
It is possible prove in the similar manner that if one moves the cut so that it passes finitely many eigenvalues of $\mathcal{H}$, the determinant does not change.
\end{remark}

\begin{remark}\label{rem:pot}
If one considers the damped wave equation with a potential term, namely,
$$
  \frac{\partial^2 v(t,x)}{\partial t^2} + 2 a(x) \frac{\partial v(t,x)}{\partial t} = \frac{\partial^2 v(t,x)}{\partial x^2} + b(x) v(t,x)\,,
$$
then it is possible to use a similar approach to the above. If the operator on the right-hand side has only negative eigenvalues,
the negative eigenvalues of the damped wave equation always appear in pairs, as in the case without the potential. The only difference is in the solution of the Cauchy problem. We define $y(x)$ satisfying $y''(x)+b(x)y(x) = 0$, $y(0)=0$, $y'(0)=1$. The result for the determinant is $\mathrm{Det\,}\mathcal{H} = 2y(T)$ for the cut of the logarithm just above the negative real axis, and $-2y(T)$ for the opposite case. If the operator on the right-hand side also has positive eigenvalues (but no zero eigenvalue), the phase $(-1)^{\mathrm{card\,}I_2}$ appears multiplying the determinant, changing the sign according to whether the number of negative eigenvalues of the damped wave equation is odd or even.
\end{remark}

\section*{Acknowledgements}
P.F. was partially supported by the Funda{\c{c}}{\~a}o para a Ci{\^e}ncia e a Tecnologia, Portugal, through project PTDC/MAT-CAL/4334/2014.
J.L. was supported by the project ``International mobilities for research activities of the University of Hradec Kr\'alov\'e'' CZ.02.2.69/0.0/0.0/16{\_}027/0008487. J.L. thanks the University of Lisbon for its hospitality during his stay in Lisbon.

\end{document}